\documentclass{article}
\usepackage[english]{babel}

\usepackage[letterpaper,top=2cm,bottom=2cm,left=3cm,right=3cm,marginparwidth=1.75cm]{geometry}
\usepackage{amsmath,amssymb,amsfonts}
\usepackage{algorithmic}
\usepackage{graphicx}
\usepackage[colorlinks=true, allcolors=blue, breaklinks=true]{hyperref}
\usepackage[square,numbers]{natbib}

\usepackage[
labelfont=sf,
hypcap=false,
format=hang,
]{caption}
\usepackage{capt-of}
\usepackage{subcaption}
\usepackage[pagewise]{lineno}

\usepackage{nomencl}
\makenomenclature

\begin{document}

\title{Displacement sensing using bi-modal resonance in over-coupled inductors}
 \author{Alexis {Hernandez Arroyo}\thanks{School of Electronic, Electrical and Mechanical Engineering, Faculty of Engineering, University of Bristol, United Kingdom, BS8 1TR}, George {Overton}*,  Anthony J. Mulholland\thanks{School of Engineering Mathematics and Technology, Faculty of Engineering, University of Bristol, United Kingdom, BS8 1TR}, Robert R. Hughes*}

\maketitle

%\linenumbers
\begin{abstract}
This paper presents the theory and key experimental findings for an investigation into the generation of bimodal resonance (frequency splitting) phenomena in mutually over-coupled inductive sensors, and its exploitation to evaluate relative separation and angular displacement between coils. This innovative measurement technique explores the bimodal resonant phenomena observed between two coil designs - solenoid and planar coil geometries.
 The proposed sensors are evaluated against first-order analytical functions and finite element models, before experimentally validating the predicted phenomenon for the different sensor configurations. The simulated and experimental results show excellent agreement and first-order best-fit functions are employed to predict displacement variables experimentally.  Co-planar separation and angular displacement are shown to be experimentally predictable to within $\pm1mm$ and $\pm1^o$ using this approach.
This study validates the first-order physics-based models employed, and demonstrates the first proof-of-principle for using resonant phenomena in inductive array sensors for evaluating relative displacement between array elements.
\end{abstract}

\section{Introduction}

Magnetic sensors are regularly used in a range of different non-contact displacement measurement applications, including in robotics~\cite{Yu2019} and have been used to measure both strain~\cite{Melzer2019} and relative displacement~\cite{Zhang2022}.
Existing technologies measure angles and displacement either by evaluating the magnetic field changes of permanent magnets using Hall-effect sensors~\cite{Tzemanaki2013HandInstrumentation,Dupre2020AAccuracy}, or the electrical properties of inductive coils~\cite{Bonfitto2019ResonantBearings, Reddy2017LowPosition}.
Real-time evaluation of surface geometries, such as local curvature, is highly desirable in many applications including structural engineering ~\cite{Chillara2020Self-sensingStructures}, medical treatment \cite{Ungi2014SpinalSnapshots}, robotics and human-computer interaction~\cite{Majidi2011AElectronics, Ward-Cherrier2018TheMorphologies}, as well as metrology~\cite{Li2012SimpleMeasurement} and non-destructive testing (NDT) of complex geometries~\cite{Mineo2017FlexibleComponents}.
Typical methods for evaluating surface curvature include the use of electrical resistance of materials, as demonstrated by Majidi et al. \cite{Majidi2011AElectronics}, where the bending curvature is determined by measuring the change of the electrical resistance of an embedded microchannel of conductive liquid.

Different approaches to non-contact measurements include the use of wireless passive LC sensors. Strongly coupled LC resonators present a frequency-splitting phenomenon - studied by Zhang et al. \cite{Zhang2014FrequencyTransfer} in power transfer applications, where the maximum efficiency is typically at the resonant frequency of the resonators. When similar resonators are positioned in close proximity, the mutual inductance between them increases significantly, resulting in a split in the resonance frequency and a decrease in the efficiency of power transfer.  However, the phenomenon of frequency splitting is not limited solely to power transfer applications.

The frequency splitting phenomenon has been extensively studied in microwave antenna theory, particularly in the context of analysing displacement in microwave sensors through frequency response~\cite{Zhu2022,Horestani2014}. 
The use of the frequency splitting phenomenon is present in different radio-frequency research studies conducted by Babu and George, they have illustrated a linear and highly sensitive displacement measurement system for wireless passive LC sensors~\cite{Babu2016,Babu2018}.
The frequency splitting phenomenon manifests when inductors exhibit a substantial mutual inductance, a property determined by the magnetic coupling coefficient, $k$, and depends on the amount of flux sharing between inductors. A such the frequency splitting phenomena is highly dependent on distance and alignment of inductors.  The resonant frequency splitting phenomenon has therefore been effectively applied to measure displacements in such sensors~\cite{Babu2019} and used across a range of different applications. Among these applications is the measurement of fluid levels inside a tank, where a passive resonator coil floats at the liquid's surface and the frequency splitting effect is measured in an external coil~\cite{Babu2019}.

The use of magnetic coupling for evaluating displacement is not limited to the use of two LC resonators. Dian Jiao et al.~\cite{Jiao2019} explored the mutual inductance between an LC resonator and conductive material, using this mutual inductance to calculate the separation and displacement between a resonator and the material.

To date, the evaluation of displacement in inductive resonant sensors has primarily focused on coaxial air-core coils. This paper seeks to extend the principles of resonant frequency splitting to over-coupled inductors in planar array configurations to evaluate relative displacement (separation and angular) in adjacent coils.  Here we exploit the magnetic coupling enhancing effects of ferrite cores and employ simple approximations to link the frequency splitting to the relevant displacement variables via the coupling coefficient.

The proof-of-principle is demonstrated in a two-coil configuration, where the evaluation of the coupling coefficient is made by monitoring the frequency spectra response in the active branch of LC sensors. The wireless and single-branch measurement allows the application of the sensors in different technologies, yet to be explored.
The different coil configuration extremes demonstrated in this paper highlight the accuracy and versatility of the modelling for designing displacement sensors.

% The evaluation of the coupling coefficient is measured in the active LC sensor, while the passive LC is floating. 
% Typically for the coupling coefficient measurement applications is necessary to connect the passive sensor, in this paper, the evaluation of the coupling coefficient, $k$, is based on the separation of the peaks in the resonance spectra~\cite{Tyurnev2010CouplingTheory}.

\label{sec:introduction}

\section{Theory}\label{sec:2}

Inductive array sensors operate on the principle of electromagnetic induction, where alternating currents generate changing magnetic fields via Ampere's law. These changing magnetic fields generated by a transmitter coil generate current in neighbouring elements \cite{Kalhor1990TheBiot-Savart}. For closely spaced coils, this effect generates mutual coupling between them. This paper focuses on the prediction, measurement, and characterization of coupling between identical resonant inductors for the evaluation of displacement in inductive sensors. This exploits a phenomenon of bimodal resonance observed by Zhang et al. \cite{Zhang2014Frequency-splittingTransfer,Babu2016} and Hughes et al. \cite{Hughes2016High-SensitivitySuperalloys} in closely packed inductors. The following subsections outline the principles behind mutual coupling and resonance in two-coil probes.

\begin{figure*}[!t]
     \centering
    \begin{tabular}{c c c} 
    a) & b) & c) \\
  \vspace{-.1cm}
   \includegraphics[width=2in]{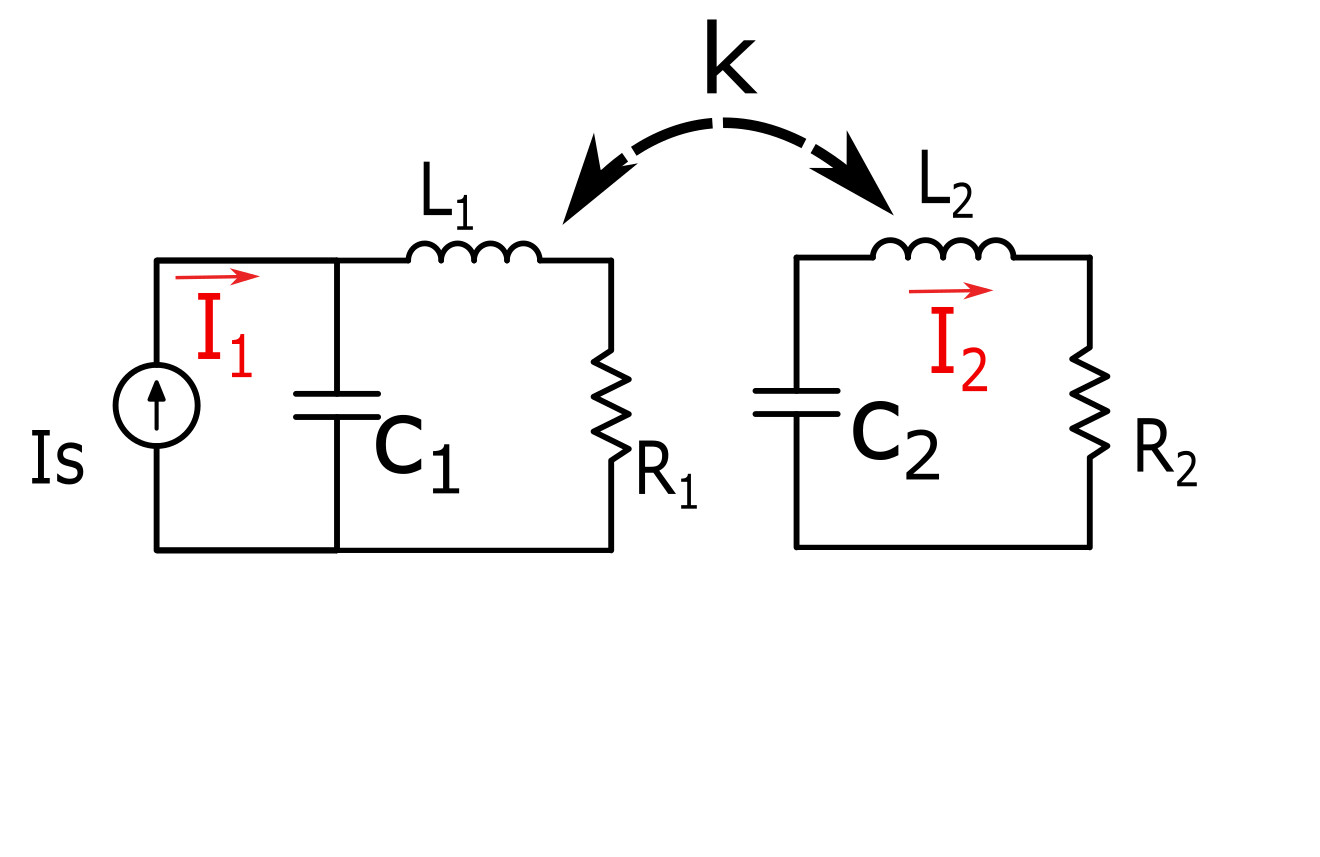}&
    \includegraphics[width=2in]{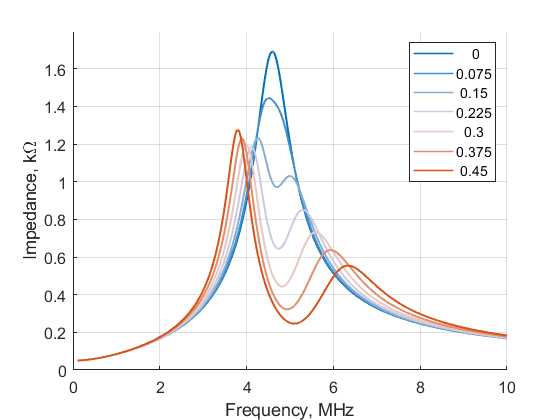}&
    \includegraphics[width=2in]{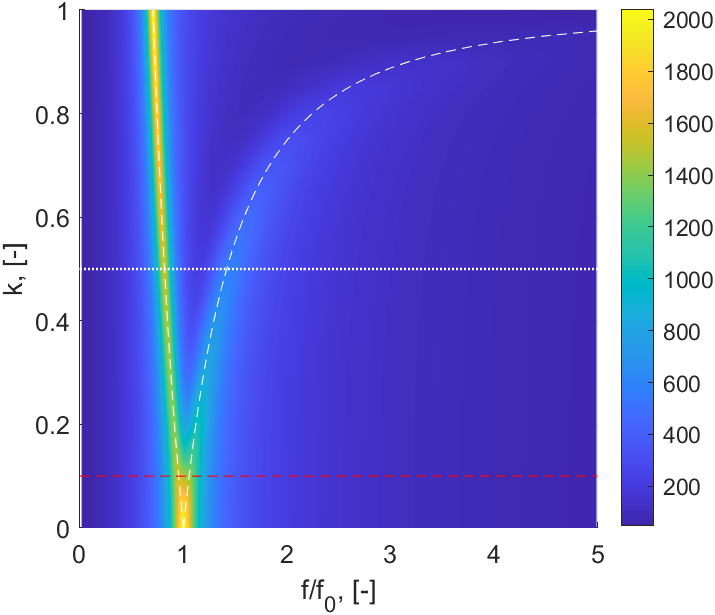}\\
 \end{tabular}
 \caption{Equivalent circuit modelled results of a two-coil over-coupled inductor system, showing; a) the equivalent circuit diagram, b) the circuit model predicted bimodal resonance phenomenon exhibited in the electrical impedance magnitude spectra as a function of the coupling coefficient, $k$, (from equation \ref{eqn:k}), and c) impedance magnitude heat map showing the trajectories of resonant peaks as a function of $k$. Circuit component values used are $R_1 = R_2 = 100\Omega$, $C_1=C_2=1nF$,  and inductance of $L_1=L_2=160\mu H$.}
    \label{fig:eqvCirc}
\end{figure*}

\subsection{Self-Resonance Equivalent Circuit Model} \label{sec2:eqvCirc}
Inductive sensors can be modelled as a parallel LC equivalent circuit (see Figure~\ref{fig:eqvCirc}.a), with the inductor, $L_1$, and the capacitor, $C_1$, representing the energy stored in the electrical field within the sensor respectively~\cite{Hughes2016High-SensitivitySuperalloys}. Typically $C_1$ is the lumped combination of capacitive effects between coil windings as well as any cabling or capacitive loads applied to the sensor. The resonant frequency, ${f}_0$, can be defined for a single inductor circuit as,

\begin{equation}
    \omega_0 = 2\pi f_0=\frac{1}{ \sqrt{LC}},\label{eqn:f0}
\end{equation}

where $\omega_0$ is the angular resonant frequency. At this natural resonance the magnitude of the impedance, $|Z|$, of the sensor is a maximum, and the voltage, $V$, and current, $I$, within the circuit are in phase (zero phase lag).

Operating inductive sensors at or near resonance have been shown by many authors to improve measurement sensitivity and maximise the mutual inductance between array elements \cite{Hughes2014NearDetection,Mazlouman2009Mid-rangeSensors, Moghaddami2018SensorlessControllers}, however, the resonant measurement techniques in array sensors have yet to be explored in detail.

\subsection{Two-Coil Mutual Resonance Model} \label{sec:2coil}
When close to another comparable coil, the primary coil will inductively couple to the secondary coil (see Figure~\ref{fig:eqvCirc}.a). The secondary coil can be modelled as an inductor, $L_2$, in series with a resistor, $R_2$, and a capacitor, $C_2$. This coupling, parameterized by the coupling coefficient, $k$, (see equation~(\ref{eqn:k_gen})) will alter the effective inductance and resistance ($L_1'$ and $R_1'$, respectively) of the primary measurement circuit and will distort the measured impedance, $Z_{m}'$, given as \cite{Hughes2014NearDetection},

\begin{align}
Z_{m}' &= \frac{R_1' + i \omega L_1'}{1+i\omega R_1' C_1 - \omega^2 L_1' C_1}.\label{eqn:Zeff1}
\end{align}
Here, $L_{1}'$ and $R_{1}'$ are given by,
\begin{align}
L_{1}' &= L_1\left[1 - \alpha^2 \frac{L_2}{L_1}\left( 1 - \frac{\omega_2^2}{\omega^{2}} \right) \right],\label{eqn:L1prime1} \\
R_{1}' &= R_1\left[1 + \alpha^2 \frac{R_2}{R_1} \right], \label{eqn:R1prime1}
\end{align}
% where $\gamma_{2} = \omega/\omega_2$ is the dimensionless frequency ratio of the secondary circuit. 
where $\alpha^2$ is defined as,
\begin{equation}
\alpha^2 = \frac{\omega^2 M^2}{R_2^2 + \omega^2 L_2^2\left(1 - \frac{\omega_2^2}{\omega^{2}} \right)^2}, \label{eqn:alpha1}
\end{equation}

\begin{figure}[!t]
\centering
\includegraphics[width=5.0in]{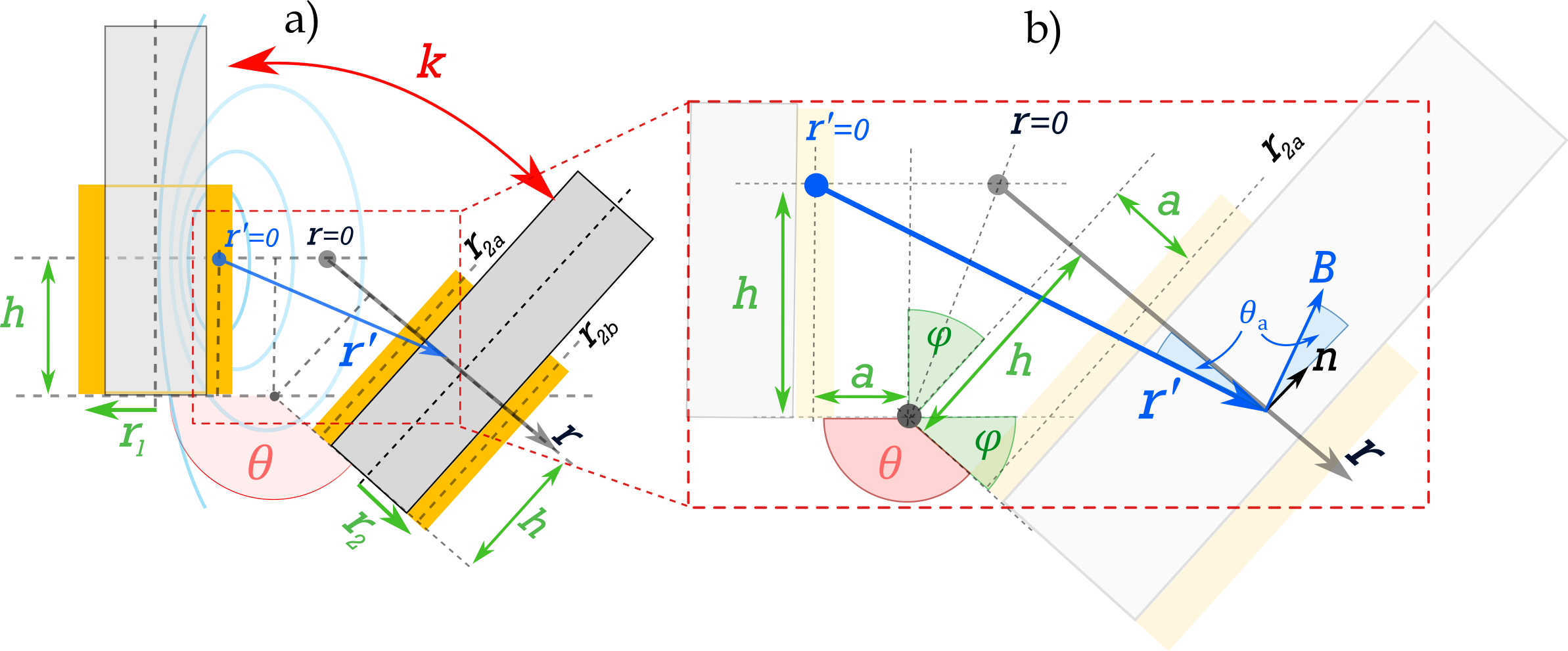}
\caption{Diagram of coupled 2D ferrite cored solenoid coils, where the yellow blocks represent the cross-section of the coil, and grey cores represent ferrite. Showing; a) angular separation $\theta$ between the centre of magnetism, $h$, of each coil, and b) a zoomed-in diagram of the trigonometric relationship between the distance from the line of symmetry ($r=0$) and the distance from the centre of the B-field source ($r'=0$), showing; separation, $a$, from the centre-point ($r=0$), and angular displacement $\phi = \pi - \theta$. Adapted from \cite{Hughes2023AnalyticalCoils}.}
\label{fig:coil-geom-big}
\end{figure}

where $M=k\sqrt{L_1L_2}$ is the mutual inductance between coils. The resonant frequencies of the two-coil system can be determined from equations~(\ref{eqn:Zeff1})-(\ref{eqn:alpha1}). The resulting expression takes the form,

\begin{align}
\omega_{\pm}' = \sqrt{ \frac{1}{2}\frac{\omega_1^2}{\left[1-k^2\right]}\left[ 1 + \frac{\omega_2^2}{\omega_1^2}\left(1 \pm \frac{\omega_1}{\omega_2} \sqrt{\frac{\omega_2^2}{\omega_1^2} + \frac{\omega_1^2}{\omega_2^2} + 4k^2 - 2} \right) \right]}, \label{eqn:w0general2c}
\end{align}

where $\omega_n = \sqrt{1/L_n C_n}$ is the uncoupled resonant frequency of coil $n$. Equation (\ref{eqn:w0general2c}) predicts two distinct resonant frequencies of the two-coil system. These can be determined for a given value of $k$ by knowing the natural resonant frequency of each coil in their uncoupled ($k=0$) state.

For the special case of identical coils where $L_1=L_2$ and $C_1=C_2$, or more specifically when their resonant frequencies are matched $\omega_1 = \omega_2$, equation (\ref{eqn:w0general2c}) simplifies to give the formula for the coupled resonant frequencies $\omega_{\pm}'$ of,

\begin{align}
\omega_{\pm}' &\approx \sqrt{ \frac{1}{C_1 L_1\left[1 \pm k \right]} }=\omega_1\sqrt{\frac{1}{1 \pm k}}.\label{eqn:f0m1}
\end{align}

Equation (\ref{eqn:f0m1}) can be used to predict the change in the resonant frequencies as a function of $k$~\cite{Babu2016}. Figure \ref{fig:eqvCirc}.b shows how the magnetic coupling $k$ affects the resonant frequency and how it produces the resonant splitting frequency phenomenon that separates the resonant peaks, shown as white dashed lines in Figure~\ref{fig:eqvCirc}.c, as a function of the coupling coefficient. The system can be thought of as exhibiting independent vibrational modes. Figure~\ref{fig:eqvCirc}.c also shows a red dotted line representing the dispersion separation threshold - the coupling coefficient of this system above which resonant peaks can be resolved as distinct peaks. This threshold is dependent on the quality factor of the systems and as such lower resistance systems exhibit sharper, more easily resolvable curves at lower coupling coefficients. There is also a practical upper threshold to the coupling coefficient of a realistic inductively coupled system which is dependent on the geometry of the system and the permeability of the cores used within the coils.

From equation (\ref{eqn:f0m1}), an expression for the coupling coefficient, $k$, between identical coils can be derived as a function of the measurable resonant frequencies observed in the spectra,
\begin{align}
k &\approx \frac{f_{-}^2 - f_{+}^2}{f_{-}^2 + f_{+}^2}\label{eqn:k}\
\approx \frac{\gamma_{\pm}^2 - 1}{\gamma_{\pm}^2 + 1}%\label{eqn:kgam}
\end{align}

where $\gamma_{\pm}={f_-}/{f_+}$ is the bimodal resonant frequency ratio.  Equation~(\ref{eqn:k}) matches the generalised Cohn-Matthaei formula for coupled resonators derived by Tyurnev 2007 \cite{Tyurnev2010CouplingTheory}.  Comparable formulas are regularly applied in the modelling of microwave bandpass networks \cite{Tung2005ACoupling} and meta-material design \cite{Petrov2015RetrievalCoefficients}, but the features and properties of coupled resonators are yet to be explored in inductive measurement applications. Rearranging equation~(\ref{eqn:f0m1}) gives an expression for the bimodal frequency ratio, $\gamma_{\pm}$, between split resonant frequencies, as a function of $k$, for an identical coil configuration,

\begin{equation}
\gamma_{\pm} = \frac{f_-}{f_+} \approx \sqrt{\frac{1 + k}{1-k}}\label{eqn:gamma}.
\end{equation}

\subsection{Coupling Coefficients} \label{sec:Ktheory}

Using the centre of magnetism (CoMag) approach detailed in  \cite{Hughes2023AnalyticalCoils}, a 2D approximation for the coupling coefficient, $k$, between neighbouring coils can be expressed as,
\begin{align}
       k = \frac{\Phi_2}{\Phi_1} &\approx \frac{1}{4\pi} \int_{r'_{2a}}^{r'_{2b}} \frac{1}{r'} \,dr' %\label{eqn:k_int}
       % &= \frac{1}{4\pi} \left[\ln{r'}\right]_{r'_{2a}}^{r'_{2b}}, \\
       \approx \frac{1}{4\pi} \ln{\left[\frac{r'_{2b}}{r'_{2a}}  \right]}, \label{eqn:k_gen}
\end{align}

where $\Phi_n$ is the magnetic flux through coil $n$, and $r'_{2a}$ and $r'_{2b}$ are defined for a coil radius, $r_2$, height, $h$, coil separation $a$, and relative angle $\theta \in$ [$\pi /2, \pi$] ($\theta = \pi - \phi$) as,

\begin{align}
     r'_{2a} &= r_2 \Lambda \sqrt{2 \left(1 -\cos \theta\right)},\label{eqn:r2a_gen}\\
     r'_{2b} &= r_2\sqrt{2\left[2 + 2\Lambda\left(1 -\cos \theta\right) + \Lambda^2 \left(1 -\cos \theta\right)\right]},\label{eqn:r2b_gen}
\end{align}
where,
\begin{equation}
\Lambda = \zeta + \eta\tan{\frac{\phi}{2}},\label{eqn:LAM}
\end{equation}

$\zeta = a/r_2$ is the separation ratio, and $\eta = h/r_2$ is the aspect ratio of the coil. Substituting the above expressions into equation~(\ref{eqn:k_gen}) allows us to calculate $k$ as a function of the separation, $a$, or relative angle $\theta$.  There are three specific cases that can be considered; co-planar separation ($\theta = \pi$), the angular displacement of planar coils ($h \ll r_2$), and angular displacement of solenoid coils ($h \gg r_2$).

\subsubsection{Co-planar Separation}\label{sec:coplanar}

When identical coils are coplanar ($\theta=\pi$), $\Lambda = \zeta$, and equations~(\ref{eqn:r2a_gen})-(\ref{eqn:r2b_gen}) can be simplified to $r'_{2a}=2a$ and $r'_{2b}=2r_2(\zeta + 1)$. We can therefore define the coupling coefficient from equation~(\ref{eqn:k_gen}) as a function of the dimensionless separation ratio, $\zeta$,
\begin{equation}
k \approx \frac{1}{4\pi} \ln{\left( \frac{\zeta + 1}{\zeta} \right)} \equiv \frac{1}{4\pi} \ln{\left( 1 + \frac{r_2}{a} \right)}. \label{eqn:kl}
\end{equation}
Equation (\ref{eqn:kl}) is valid for all coil aspect ratios. Equation~\ref{eqn:kl} can be rearranged to arrive at a generalised linear function of the form $y = p_1 x + p_2$,
\begin{equation}
e^{4\pi k} \approx p_1\frac{1}{a} + p_2, \label{eqn:k2}
\end{equation}
where $p_1$ and $p_2$ are unknown coefficients of the first order polynomial, and can be fitted to experimental data to enable displacement prediction \cite{Hughes2023AnalyticalCoils}. 
\par
\vspace*{2mm}

\subsubsection{Angular Displacement - Planar coils}
Consider two planar coils with a varying angle as is shown in Figure \ref{fig:coil-geom-big}, with $h \ll r_2$ such that the coil aspect ratio can be considered negligible ($\eta \ll \zeta$) coils are planar. In this case for an angle $\phi$ away from co-planar (i.e. $\theta < \pi$), $\Lambda \approx \zeta$ such that equations~(\ref{eqn:r2a_gen})-(\ref{eqn:r2b_gen}) can be simplified to express the coupling coefficient as follows \cite{Hughes2023AnalyticalCoils},
\begin{align}
       k \approx \frac{1}{8\pi} \ln{\left[1 + \frac{2}{\zeta}+\frac{4}{\zeta^2\left(4 - \phi^2\right)}\right]},\label{eqn:k_ang_small}
\end{align}
where $\phi$ is given in radians. 
Equation~\ref{eqn:k_ang_small} can be simplified into a generalised linear function for the angular displacement, $\phi$, for filament coils,
\begin{align}
        e^{8\pi k} \approx p_1\frac{1}{\left(4 - \phi^2\right)} + p_2, \label{eqn:phi_ang_small}
\end{align}
where again $p_1$ and $p_2$ are the unknown coefficients of a first order polynomial, which can be found by fitting to experimental data \cite{Hughes2023AnalyticalCoils}.
\par
\vspace*{2mm}

\subsubsection{Angular Displacement - Solenoid coils}
For the case when $a \ll h$, $\zeta \ll \eta$, a first-order approximation can be derived for $k$ as \cite{Hughes2023AnalyticalCoils},
\begin{equation}
    k \approx \frac{1}{4\pi} \ln{\left[ 1 + \frac{2}{\eta\phi} \right]}.
    \label{ksolenoid}
\end{equation}
As before, a generalised linear function for $\phi$ can be defined as,
\begin{equation}
    \eta e^{4\pi k} \approx p_1\frac{1}{\phi} + p_2. \label{eqn:phi_ang_big}
\end{equation}
% The predicted behaviours demonstrated in Figure~\ref{fig:ksepRat} were evaluated experimentally for the two scenarios; 2D planar (PCB), and solenoid coils.  
Equations~(\ref{eqn:kl}) and (\ref{ksolenoid}) provide simple functions with which to predict $\phi$ between coils when $k$ is experimentally measurable. While the extreme approximations used to arrive at these functions means they may not provide accurate absolute values, their generalised functions given in equations~(\ref{eqn:phi_ang_small}) and (\ref{eqn:phi_ang_big}) can be used to fit experimental data \cite{Hughes2023AnalyticalCoils}. 

\section{Materials \& Methods}\label{sec:3}
Finite element modelling (FEM) and experimental tests were conducted to validate and characterise the bimodal resonance behaviour predicted by the circuit theory (Section~\ref{sec:2coil}). Complete details of the modelling and experimental methods are detailed in the following subsections.

 \begin{figure*}[!t]
    \centering
    \includegraphics[width=6in]{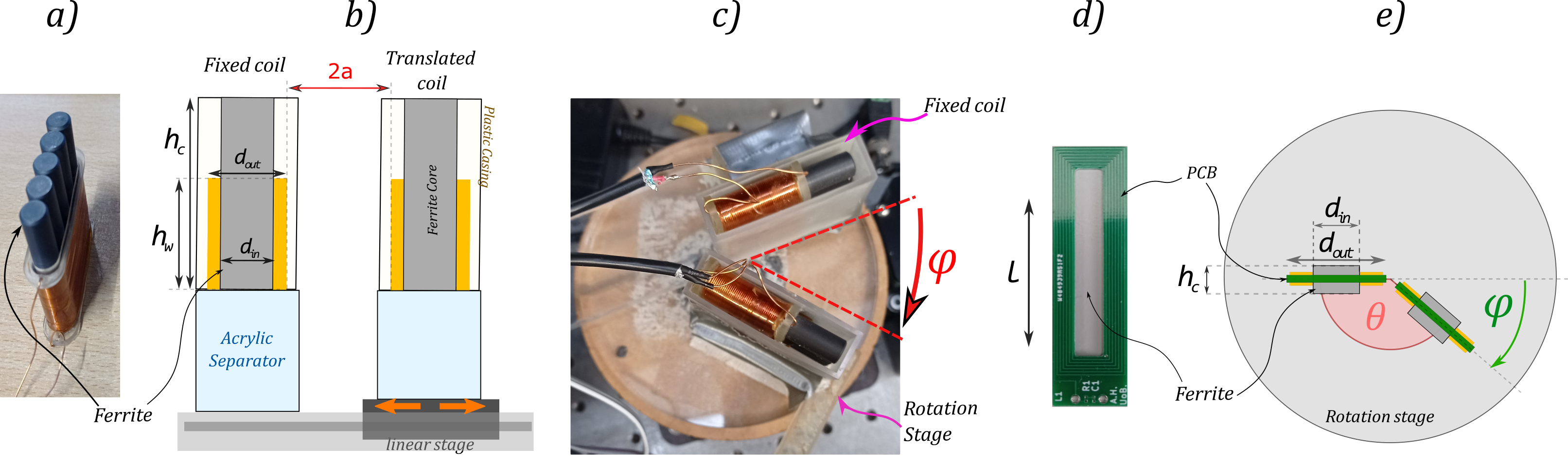}
         \caption{Coil design and experimental setup for coplanar separation and angular displacement studies. Showing a) Solenoid ferrite cored coil top-down design, along with b) cross-sectional coplanar separation, $a$, and c) angular displacement, $\phi$, testing configurations. d) PCB-type (planar) coil top-down design and e) angular displacement, experimental configuration.} \label{fig:Experimental setup}
\label{sep_ang_all}
\end{figure*}

\subsection{Sensor Designs}\label{SensorDesign}

To test the equations previously defined two different extremes of coils were developed, with two different studies: co-planar separation and angular displacement.

For construction simplicity, the solenoid coil was designed with a single layer of forty turns with a wire diameter of $0.56~mm$, and the planar coil was a double-layer printed circuit board (PCB) with a rectangular shape (see Figure \ref{fig:Experimental setup}.d). The electrical and geometrical parameters of the solenoid and planar coils are shown in Table~\ref{table}.  Experimental coil dimensions are given with an error of $\pm0.5~mm$, and experimental circuit parameters are given with tolerances of $\pm 5\%$.

\begin{table}[!b]
\centering
\captionof{table}{Experimental (Exp.) and simulated sensor coil dimensions and electrical circuit parameters for solenoid (Sol.) and planar (Plan.) coils in experimental tests, as well as 2D and 3D finite element simulations}.
\vspace{1mm}
%\centering
\setlength{\tabcolsep}{3pt}
\begin{tabular}{l|c c c c c}
 \multicolumn{6}{c}{\textbf{Coil Dimensions(mm)}} \\
 \hline
 Coil Type & $l$ & $h_c$ & $h_w$ & $d_{in}$ & $d_{out}$\\
\hline
Sol & 43 & 40 & 25 & 8 & 11 \\
 Plan. & 20 & 1.5 & 1.0 & 5 & 12 \\
 \hline 
 \multicolumn{6}{c}{} \\
  \multicolumn{6}{c}{\textbf{Circuit Parameters}} \\
 \hline
Coil Type & Turns & $L$ ($\mu H$) & $C$ ($nF$) & $R$ ($k\Omega$) & $I$ ($mA$)\\
\hline
Sol. (2D) & 40 & 708 & 0.2 & 0.1 &  10 \\
Sol. (3D) & 40 & 170 & 1 & 1 & 10 \\
Sol. (exp.)  & 40 & 166 & 1 & 1 & 10 \\
\hline
Plan. (2D)  & 28 & 7.9 & 0.9 & 0.01 & 1 \\
Plan. (3D)  & 28 & 7.5 & 1 & 0.01 & 1 \\
Plan. (exp.)  & 28 & 6.7 &  1 & 0.01 & 1 \\
 \hline

 \end{tabular}
\label{table}
\end{table}

The inductance of the coil was obtained using an impedance analyser TREWMAC TE3001 (TrewMac Sys-tems, Australia). For the 2D model and 3D model, geometry coil analysis module in COMSOL was used to obtain the values of the inductances, the capacitor and resistor used the experimental values obtained previously.

 The coils are connected to a resonant tank circuit \cite{Wang2018AnTransfer}; commonly used for radio frequency and also in signal filtering applications. The solenoid coils were fabricated in-house, while the planar coils were manufactured by PCBway, China.

\subsection{Finite Element Modelling}\label{FEM}
Parameterised finite element models (FEM) were created in COMSOL 6.1 using the AC/DC module. While this paper is not focused on the optimization of the magnetic field in inductors, the magnetic coupling of the inductors is directly related to the generation of the bimodal phenomenon. Models were developed using Magnetic Fields and Electrical Circuit studies to simulate the bimodal phenomenon. These are compared to experimental measurements of two extreme coil geometries, planar and solenoid coils.

\begin{figure}[!t]
    \centering
    \includegraphics[width=3.5in]{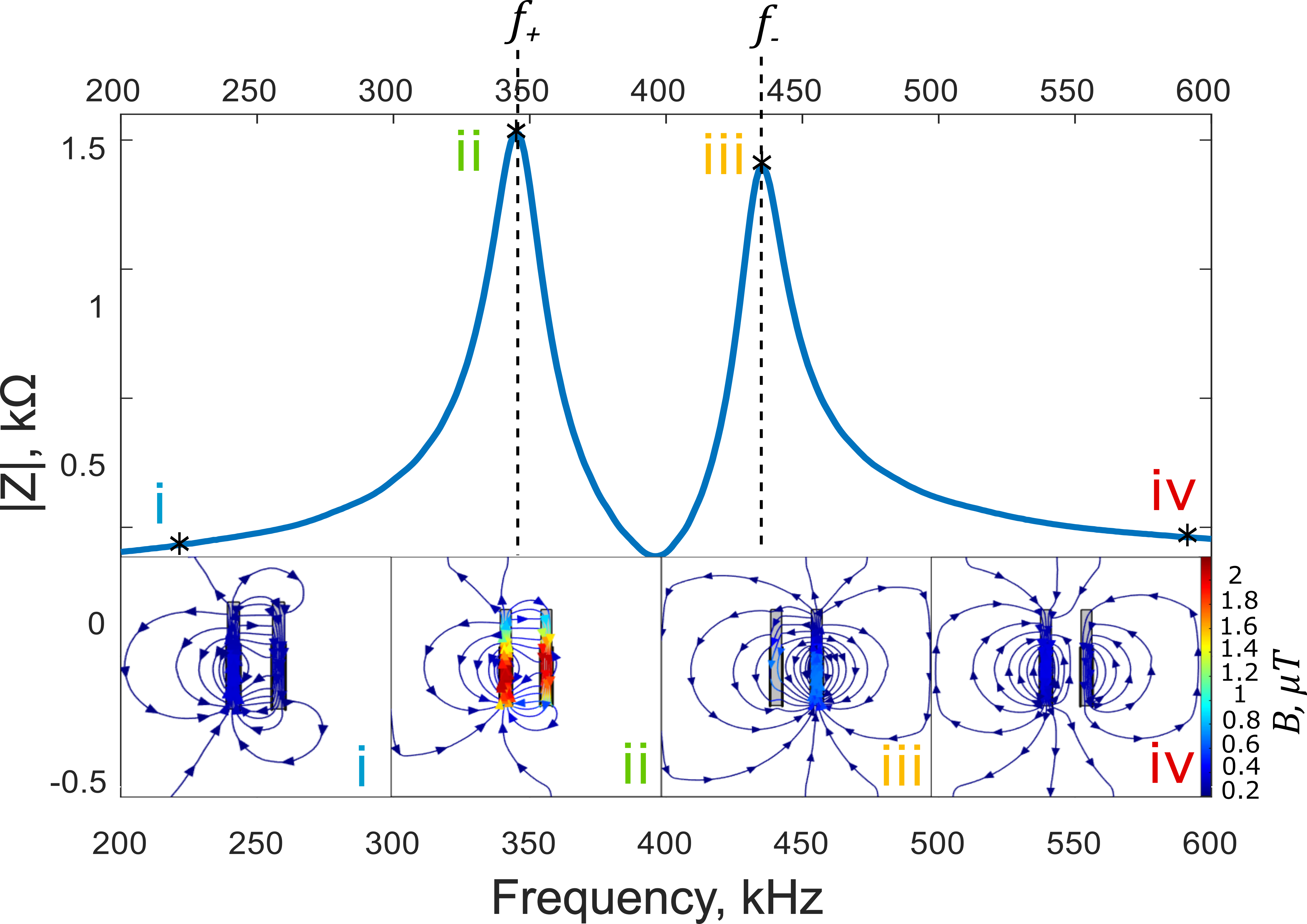}
    \caption{Finitie element simulation of the magnetic flux density, $B$, in a bimodal resonant system in relation to the electrical impedance magnitude $|Z|$ as a function of frequency, $f$. The spatial distribution of the in-plane $B$-field is shown at; i) 225 kHz, ii) 350 kHz (first resonant peak, $f_+$), iii) 450 kHz (second resonant peak, $f_-$), and iv) 590 kHz.}
    \label{fig:mffrequency}
\end{figure}

The parameters used for the modelled bimodal sensors are shown in Table~\ref{table}.  Figure~\ref{fig:mffrequency} shows an example simulated frequency spectrum and how the magnetic flux density, $B$, changes within an over-coupled two-coil system as a function of frequency, $f$. It is evident that at the first resonant peak ($f=f_+$) the magnetic flux is predominantly emitted from the driver coil (Figure~\ref{fig:mffrequency}.ii), while when $f=f_-$, the flux is emitted predominantly from the passive coil (Figure~\ref{fig:mffrequency}.iii).

The magnetic flux density shown in Figure~\ref{fig:mffrequency}.ii is maximum at the first resonance peak, which corresponds to a maximum $B$-field in both the driver and passive coil, generating maximum power transfer. This effect of maximum power transfer has already been explored by Zhang et al. \cite{Zhang2015SelectiveFrequencies}.

Figure~\ref{fig:mffrequency}.i shows how the $B$-field is concentrated around the driver coil (left) for $f<f_+$ of the coils. Figure~\ref{fig:mffrequency}.ii shows that at $f=f_+$, $B$ increases significantly for both coils but the spatial distribution remains comparable. Figure~\ref{fig:mffrequency}.iii shows that $B$ is concentrated in the passive coil at $f=f_-$, and Figure~\ref{fig:mffrequency}.iv shows how $B$ is distributed evenly at higher frequencies beyond the resonant peaks.

\subsection{Experimental Measurements} 

The two sensor configurations defined in Table~\ref{table} had their bimodal impedance spectra evaluated experimentally as a function of their coplanar separation, $a$, and angular displacement, $\phi$. The equipment and methods used are summarised below.

\vspace*{2mm}

\subsubsection{Data Acquisition}\label{dataAq}
 A TREWMAC TE3001 (TrewMac Systems, Australia) impedance analyser was used to measure the impedance across the driver coil. The impedance analyzer took a total of 1024 points in a frequency sweep between 0.3-0.5 MHz for the solenoid coils and between 1.7-2.2 MHz for the planar coils. This data was recorded and, using an automated peak detection process, the frequencies of the bimodal resonant peals were determined. 
For each angle, a frequency sweep is recorded using the impedance analyzer, and the relative differences between the bimodal resonance peaks are determined.
As shown in Figure \ref{sep_ang_all}.b, the coils are contained in a plastic fixture of a wall thickness of 1.5 $mm$ which generates an initial separation between the coils.
\vspace*{2mm}

\subsubsection{Co-planar Separation,  \texorpdfstring{$a$}{}} \label{Separation}

An X-LSM200A-E03 (Zaber, Canada) linear stage was attached to the passive sensor (see Figure~\ref{sep_ang_all}.b), and moved in increments of $10~mm$, between a coil centre-to-centre distance range of $15-75~mm$ (beyond which the coupling is too low for the resonance frequency to be observed in the resonance frequency sweep).
A second measurement study was conducted with separation increments of $1~mm$ to provide a higher resolution dataset used to predict the separation between the sensors.
\vspace*{2mm}

\subsubsection{Angular Displacement, \texorpdfstring{$\phi$}{}}\label{Angle}

An X-RSB120AU rotational stage (Zaber, Canada) was used to incrementally rotate a passive resonator coil relative to a fixed driver coil, with the centre of rotation at the vertex of the shared corner as shown in Figure~\ref{sep_ang_all}.c for solenoid coils and in Figure \ref{sep_ang_all}.e for planar coils.  Two sets of measurements were conducted: firstly $\phi$, was varied between $0-70^o$ in $10^o$ increments, the resonant frequencies recorded and a best-fit function (derived from the formula in section \ref{sec:Ktheory}) applied to the data. The resonant frequencies were measured again, this time for $\phi$ in increments of $1^o$, and the best-fit functions (see section~\ref{sec:Ktheory}) used to invert $\phi$.

\section{Results and discussion}\label{sec:4}
Experimental and FE results are compared directly for each of the studies, exploring co-planar separation, $a$, and angular displacement, $\phi$. The results show findings varying the coplanar separation, $a$, for solenoid coils, and $\phi$ for both solenoid and planar coil configurations.
The results are compared to best-fit functions defined in section \ref{sec:2}.

\begin{figure*}[!t]
\centering
\includegraphics[width=6in]{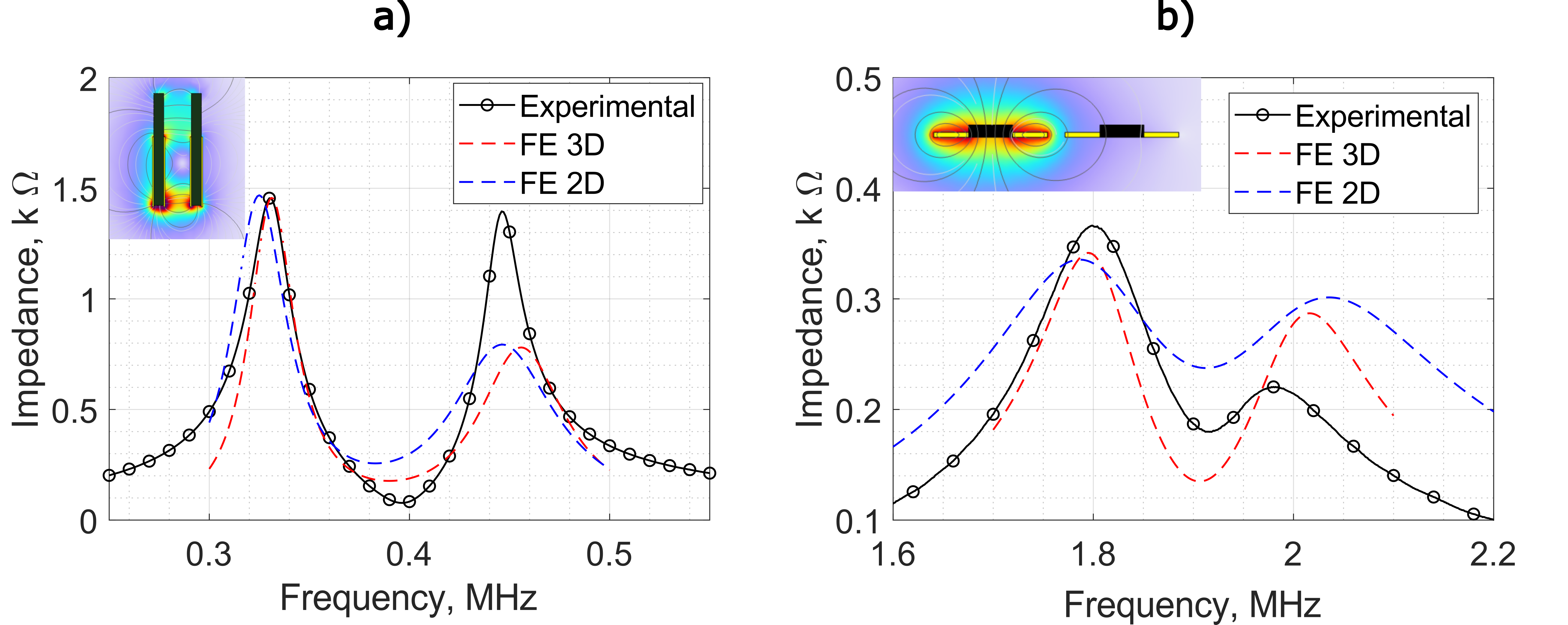}
 \caption{Comparison between simulated and experimental electrical resonant impedance spectra in an over-coupled driver coil. Graphs compare experimental results to 2D and 3D finite element modelled results (blue and red dashed curves respectively) for a) solenoid coils, and b) planar coils.}
     \label{Bimodalall}
\end{figure*}

The experimental impedance spectra of the bimodal sensors for those two coil geometries are compared to the FE models in Figure~\ref{Bimodalall}. The relationship between bimodal resonant frequencies and displacements $a$ and $\phi$ is then explored.

Equation~(\ref{eqn:k}) is used to calculate $k$ from measurable resonant frequencies, $f_-$ and $f_+$, and then the first-order approximations (section~\ref{sec:Ktheory}) can be used to estimate displacement variables.

\subsection{FE Model Validation}
The experimental impedance spectra results were compared to FE simulations in 2D and 3D and show differences in the values of capacitance as shown in Table \ref{table}, due to the separation of the wire in the hand-wound coils. These changes in capacitance lead to a different resonant frequency.
To match the resonant frequencies, it is necessary to introduce a calibration value of capacitance for the resonators. This allows the simulated system to resonate at the same frequencies as the experiment.

For the solenoid and planar coils, the resonant frequencies for the three different results are approximately matched in frequency value. The magnitude depends on the impedance of the system, for the solenoid coil, there is a better approximation of magnitude than for the planar coil. Figure \ref{Bimodalall}.a, shows the frequency spectrum for the solenoid coils with an initial centre-to-centre separation of $1.5~mm$; this separation $a$ represents the separation between the edges of the windings in the coils (see Figure~\ref{fig:coil-geom-big}).

\begin{figure*}[!t]
    \centering
    \includegraphics[width=6in]{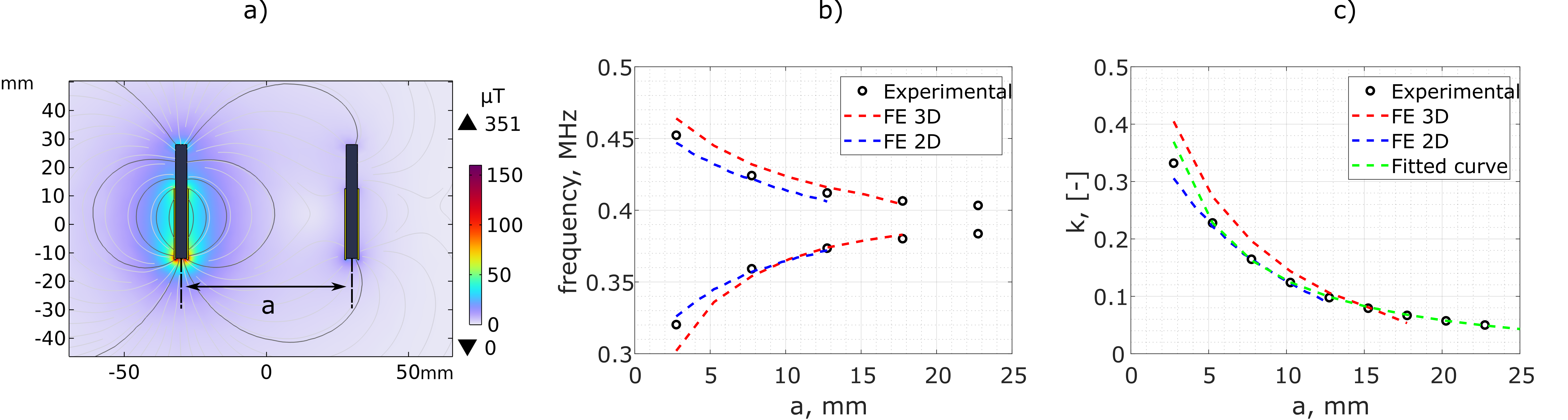}
\caption{Bimodal resonant frequencies as a function of coplanar separation, $a$, showing: a) 2D finite element modelled magnetic flux density and field lines between bimodal sensor elements for a separation of $a=60~mm$ in solenoid coils. b) Bimodal resonant frequencies, $f_-$ (lower curves) and $f_+$ (upper curves), of the system as a function of angular displacement, comparing experimental (black circles) to 2D (blue dashed) and 3D (red dashed) finite element simulated results. iii) The coupling coefficient, $k$, calculated using measurements of $f_-$ and $f_+$ from equation~(\ref{eqn:k}), where the green dashed line represents the experimental best fit function, as defined by equation~(\ref{eqn:k2}) in section~(\ref{sec:Ktheory}).}
        \label{Hdis}
\end{figure*}

For an initial separation of $a=3~mm$ between the solenoid windings due to the plastic case, the solenoid coils produce the initial bimodal resonance frequency spectrum shown in Figure \ref{Bimodalall}.a. Planar PCB coils, with an initial separation of $a=2~mm$, produce the bimodal frequency spectrum shown in Figure \ref{Bimodalall}.b. When the coil decreases in size, the number of turns reduces, making it harder to match the finite element simulation with the experimental data due to the homogeneous multiturn approximation. Despite the differences in the impedance frequency spectrum, the resonant frequencies exhibit similar trends as a function of changes in coupling as shown in Figure~\ref{Hdis}.c. 

\subsection{Co-planar Separation, \texorpdfstring{$a$}{}}\label{sepexperimental}

The bimodal resonant frequencies $f_-$ and $f_+$ were used to determine $k$ from equation~(\ref{eqn:k}) as a function of $a$. As $a$ increases, $k$ decreases causing $f_-$ and $f_+$ to converge in the frequency spectrum until they merge, as demonstrated in Figure~\ref{fig:eqvCirc}.b.

For the 2D and 3D FE, the bimodal frequency spectrum merges into a single frequency at separations greater than $12.5~mm$ and $17.5~mm$ respectively, where the passive coil has minimal interaction with the driver coil. This phenomenon has been explored by Zhang et al. \cite{Zhang2014FrequencyTransfer} for wireless power transfer applications.

\begin{figure*}[!t]
    \centering
    \includegraphics[width=6in]{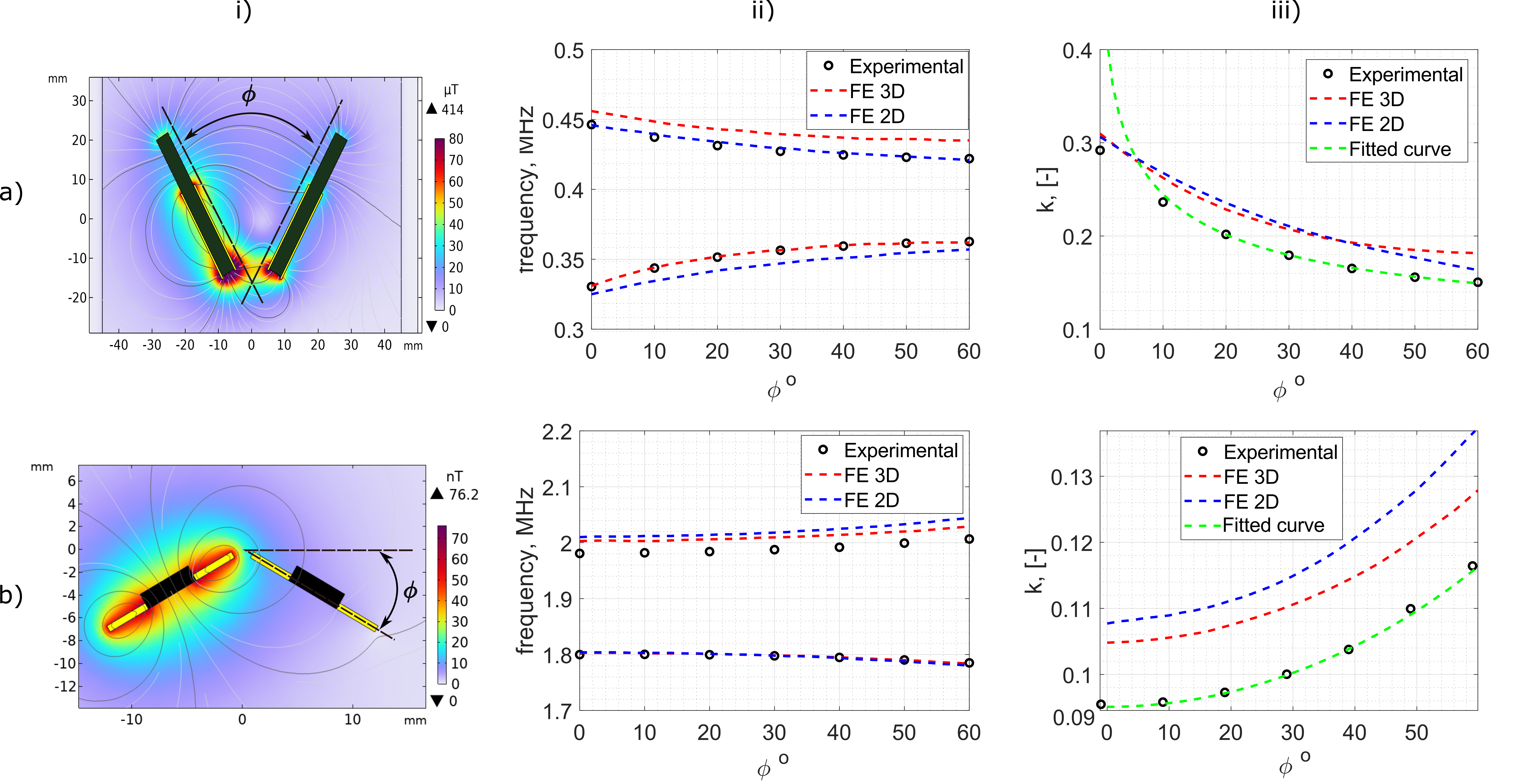}
    \caption{Bimodal resonant frequencies as a function of angular displacement, $\phi$, in a) solenoid and b) planar coils, showing; i) 2D finite element modelled magnetic flux density and field lines between bimodal sensor elements for a separation of $\phi=60^o$ in solenoid coils. ii) Bimodal resonant frequencies, $f_+$ (lower curves) and $f_-$ (upper curves), of the system as a function of angular displacement, comparing experimental (black circles) to 2D (blue dashed) and 3D (red dashed) finite element simulated results. iii) The coupling coefficient, $k$, calculated using measurements of $f_-$ and $f_+$ from equation~(\ref{eqn:k}), where the green dashed lines represents the experimental best fit function, as defined by equations~(\ref{eqn:phi_ang_big}) and (\ref{eqn:phi_ang_small}) in section~\ref{sec:Ktheory}. }
    \label{Adis}
\end{figure*}

\subsection{Angular Displacement, \texorpdfstring{$\phi$}{}}\label{angle}

The results in Figure~\ref{Adis}.a.ii show how the resonant frequencies $f_+$ and $f_-$ get closer together as a function of $\phi$ for the solenoid coils, showing comparable results to the evaluation of coplanar separation, $a$. In this case, $k$ decreases as $\phi$ increases.
Figure~\ref{Adis}.b shows  $f_+$ and $f_-$, as a function of $\phi$ for the planar coils. In this case, $k$ increases as $\phi$ increases.

The $k$ between the solenoid coils varies between $0.15-0.35$ as shown in Figure~\ref{Adis}.a.iii. This is relatively high for non-coaxial coils, where a typical value is $k<0.1$~\cite{Yang2017DesignCoils}. These high $k$ values are due to the large ferrite cores and the significant magnetic bridging it causes between the coils. The planar coils exhibit more modest values of $k$, between $0.9-0.14$. As $k$ between the planar sensors is relatively low compared to the solenoid coils (see Figure~\ref{Adis}.b.iii), the matching between the FE models and experiment becomes more challenging, leading to an offset between results for $k$ of the planar coil geometry. 
 
 The angular separation of the ferrites cores in solenoid coils increases the distance between the centres of magnetism of the cores (see section~\ref{sec:Ktheory}), decreasing $k$. Conversely, the distance between the centres of magnetism for the planar coils reduces as $\phi$ increases, causing an increase in $k$. These trends are clearly shown in Figure~\ref{Adis}.a.iii and b.iii. Figure~\ref{Adis} therefore shows excellent agreement between the experimental and FE simulations, with $f_+$ and $f_-$ changing as predicted in section~\ref{sec:Ktheory}.

\subsection{Displacement Prediction}

The experimental displacement measurements shown in Figures~\ref{Hdis} and \ref{Adis} for coplanar separations, $a$,  and angular displacements, $\phi$, with increments of $\Delta a=10~mm$ and $\Delta \phi = 10^o$ respectively, were used to fit the first-order functions defined in section~\ref{sec:Ktheory}. 

From the values of $k$ calculated using equation~(\ref{eqn:k}), linear functions of the form $y=p_1 x + p_2$ were plotted based on equations~(\ref{eqn:k2}), (\ref{eqn:phi_ang_small}) and (\ref{eqn:phi_ang_big}), as defined in Table~\ref{table2} and \cite{Hughes2023AnalyticalCoils}.
Linear regression was then used to calculate the coefficients $p_1$ and $p_2$ shown in Table~\ref{table2}. Note that the factor of $4$ in equation~(\ref{eqn:k2}) was omitted as this gave rise to a stronger linear correlation as found by \cite{Hughes2023AnalyticalCoils}. This disparity is likely due to the simplicity of the assumptions employed in the derivations of the first-order functions derived in \cite{Hughes2023AnalyticalCoils}.

The best-fit functions defined by Table~\ref{table2} were rearranged to produce the best-fit curves (green dashed) for $k$ as a function of the displacement variables as shown in Figures~\ref{Hdis}.c and ~\ref{Adis}.iii. The best-fit curves based on these first-order approximate functions exhibit excellent agreement with the experimental data over a wide range of displacements in all cases. The only exception occurring when $\phi<10^o$ in the solenoid coils, where the assumptions and simplifications of the first-order model break down. 
Studies were conducted to evaluate the effectiveness of using these functions to predict the physical displacement between sensors based on the experimental measurements of the bimodal resonant frequencies ($f_-$, $f_+$).  
\vspace*{2mm}

\begin{figure*} [t!]
\begin{tabular}{cccc}
  & 
 \hspace{-60mm} a)& 
 &
\hspace{-60mm} b)\\
 \hspace*{-1.5cm}
 i)& 
 \hspace*{-1.5cm}
 ii)&
 \hspace*{-1.5cm}
 i)&
 \hspace*{-1.5cm}
 ii)\\
 \hspace*{-1.5cm}
 \includegraphics[width=2in]{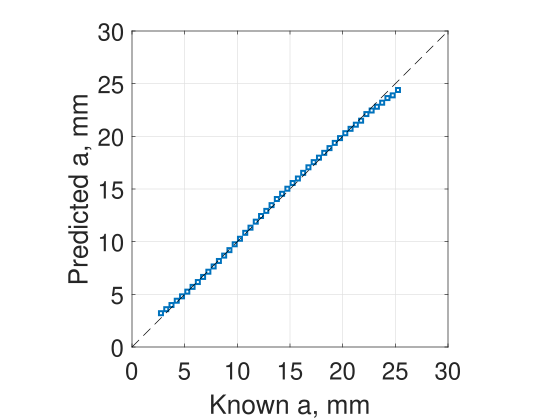}&
   \hspace*{-1.4cm}
\includegraphics[width=2in]{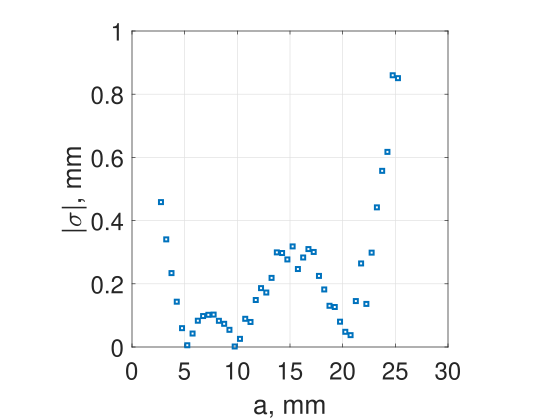}&
   \hspace*{-1.4cm}
 \includegraphics[width=2in]{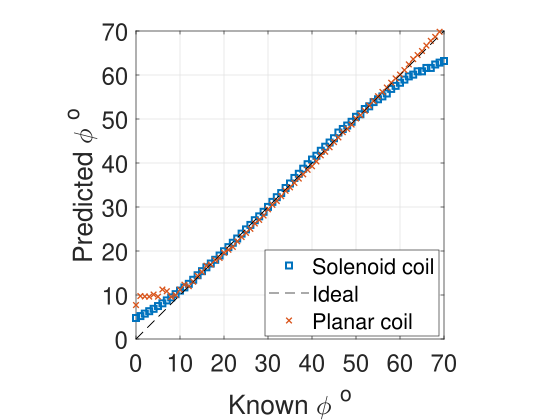}&
  \hspace*{-1.4cm}
 \includegraphics[width=2in]{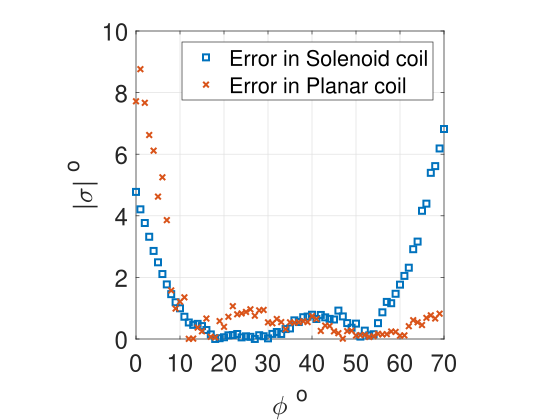}\\
\end{tabular}

    \caption{Displacement predictions of bimodal sensors, for a) coplanar separation, $a$, and b) angular displacement, $\phi$, between sensor coils. Graphs i) plot the predicted displacement (calculated from experimental measurements and using the fitted functions in Table~\ref{table2}) against the known displacement values. Black dashed lines represent ideal prediction. Graphs ii) show the accuracy of the predictions by plotting the absolute error $|\sigma|$ between predicted and known values as a function of displacement.}
    \label{Error}
\end{figure*}

\subsubsection{Evaluating Separation, \texorpdfstring{$a$}{}}\label{separation}

Experimental measurements, made every $\Delta a=1~mm$, were used and $f_+$ and $f_-$ were recorded and used to calculate the $a$ using equation~(\ref{eqn:k}) and the best-fit function defined in Table~\ref{table2}.  The prediction of $a$ has a limit of $a=25~mm$ beyond which $k$ becomes negligible and $f_+$ and $f_-$ are no longer detectable, as demonstrated in Figure~\ref{fig:eqvCirc}.b \& c.
\vspace*{2mm}

\subsubsection{Evaluating Angle, \texorpdfstring{$\phi$}{}}\label{Curvature}

The $f_+$ and $f_-$ peaks were recorded for variations in $\phi$ in both solenoid and planar coil sensors every $\Delta \phi=1^o$, and the best-fit functions (Table~\ref{table2}) were used to predict $\phi$, the results are shown in Fig~\ref{Error}.b.i.

\begin{table}[h!]
\centering
\captionof{table}{ Linear regression of the function $y=p_1 x + p_2$ for the displacement of the coils as shown in Figure~(\ref{Hdis}) and (\ref{Adis}) }
\vspace{2mm}
%    \centering
\setlength{\tabcolsep}{4pt}
\begin{tabular}{l l c c c}
%\begin{tabular}{ c l c l c c c }
     \textbf{Displacement} & $y$ & $x$ & $p_1$ & $p_2$ \\
    \hline
    $a$ & $e^{\pi k}$ & $1/a$ &  0.39 & -0.025\\
    \noalign{\smallskip}
    %\vspace{2mm}
    %\hline\\
    % \noalign{\smallskip}
    \hline
    % \noalign{\smallskip}
    $\phi$ (sol.) & $e^{8\pi k}$ & $1/\phi$ & 3.98 & 3.75\\
    \noalign{\smallskip}
    $\phi$ (plan.) & $e^{4\pi k}$ & $1/(4-\phi^2)$ & 81.83 & -9.40\\
    \hline
    \end{tabular}
    \label{table2}
\end{table}

%\vspace{-10mm}
\vspace*{2mm}

\subsubsection{Prediction Error,\texorpdfstring{ $|\sigma|$}{}}

Figure~\ref{Error}.a.i compares the predicted coil separation, $a$, using the fitted function in Table~\ref{table2} compared to that recorded experimentally from the linear translation stage. The absolute error, $|\sigma|$, between the known and predicted values of $a$ is shown in Figure~\ref{Error}.a.ii.
The results show that the measurement is accurate to within $\pm1~mm$ over the full range of distances measured.

For both solenoid and planar coils, the predictions are more accurate for $\phi>10^o$. At low angles ($\phi<10^o$), the planar sensors do not experience a significant change in $k$ (see Figure~\ref{Adis}.b.iii). Hence the error in the prediction of $\phi$ is larger for small angles. Conversely, $k$ changes significantly at low angles for the solenoid coils, however, the function employed to fit the solenoid angular displacement has low accuracy at low angles due to the assumptions employed. As such the predictions based on this function are also less accurate when $\phi<10^o$.

As shown in Figure~\ref{Error}.b.ii, the absolute error in angle for planar sensors decreases to less than $\pm1^o$ for angles $\phi>10^o$. Despite the coil unoptimised design, the planar coils exhibit excellent prediction capability up between $10^o - 70^o$.

\section{Conclusions}\label{sec:5}

Pairs of over-coupled inductive coils were used to produce bimodal resonance (frequency splitting) spectra, and resonant frequency tracking was used to evaluate the separation and angular displacement between the coils. The study demonstrated the displacement dependent bimodal phenomenon in two coil designs experimentally, and highlighted the validity of a "centre-of-magnetism" method for defining first-order analytical functions of the coupling interactions between coils. 2D and 3D FE simulations were evaluated, showing excellent agreement in with experimental results, and direct inversion of displacement variables showed the technique was able to resolve separations and angles to within $\pm1~mm$ and $\pm1^o$. 

While these resolutions are an order of magnitude larger than hall-effect techniques, the proof-of-principle in these unoptimised sensor designs demonstrates significant potential for further development of this novel sensing approach.  The bimodal resonance tracking method proposed, has the potential to be made real-time, deployed directly onto flexible substrates, and expanded to higher degrees-of-freedom making it valuable in applications such as robotics, human-computer interaction, and non-destructive testing.

\section*{Acknowledgements}

The work presented relates to a pending UK patent application - Inductive Coil Array - UK patent application no. 2215187.2, filed on 14th October 2022.  Alexis Hernandez's research is funded by the Consejo Nacional de Ciencia y Tecnología (CONACYT).  Data associated with the study will be made available via the Bristol Research Data repository upon acceptance.

\bibliography{references}

\begin{thebibliography}{10}

\bibitem{Yu2019}
Xiaojie Yu, Qiao Hu, Dan Xu, Peng Li, and Xiangpeng Liu.
\newblock Common sensors in industrial robots: A review.
\newblock {\em Journal of Physics: Conference Series}, 1267:012036, 7 2019.

\bibitem{Melzer2019}
M.~Melzer, D.~Makarov, and O.~G. Schmidt.
\newblock A review on stretchable magnetic field sensorics.
\newblock {\em Journal of Physics D: Applied Physics}, 53:083002, 12 2019.

\bibitem{Zhang2022}
Zhang J.
\newblock A displacement sensing method based on permanent magnet and magnetic flux measurement.
\newblock {\em Sensors 2022, Vol. 22, Page 4326}, 22:4326, 6 2022.

\bibitem{Tzemanaki2013HandInstrumentation}
Antonia Tzemanaki, Xinge Gao, Anthony~G. Pipe, Chris Melhuish, and Sanja Dogramadzi.
\newblock {Hand exoskeleton for remote control of minimally invasive surgical anthropomorphic instrumentation}, 6 2013.

\bibitem{Dupre2020AAccuracy}
Nicolas Dupre, Yves Bidaux, Olivier Dubrulle, and Gael~F. Close.
\newblock {A Stray-Field-Immune Magnetic Displacement Sensor with 1{\%} Accuracy}.
\newblock {\em IEEE Sensors Journal}, 20(19):11405--11411, 10 2020.

\bibitem{Bonfitto2019ResonantBearings}
Angelo Bonfitto, Ran Gabai, Andrea Tonoli, Luis~Miguel Castellanos, and Nicola Amati.
\newblock {Resonant inductive displacement sensor for active magnetic bearings}.
\newblock {\em Sensors and Actuators A: Physical}, 287:84--92, 3 2019.

\bibitem{Reddy2017LowPosition}
Battu~Prakash Reddy, Ashwin Murali, and Ganesh Shaga.
\newblock {Low cost planar coil structure for inductive sensors to measure absolute angular position}.
\newblock {\em Proceedings of 2017 2nd International Conference on Frontiers of Sensors Technologies, ICFST 2017}, 2017-January:14--18, 12 2017.

\bibitem{Chillara2020Self-sensingStructures}
V.~S.C. Chillara, A.~K. Ramanathan, and M.~J. Dapino.
\newblock {Self-sensing piezoelectric bistable laminates for morphing structures}.
\newblock {\em Smart Materials and Structures}, 29(8):085008, 6 2020.

\bibitem{Ungi2014SpinalSnapshots}
Tamas Ungi, Franklin King, Michael Kempston, Zsuzsanna Keri, Andras Lasso, Parvin Mousavi, John Rudan, Daniel~P. Borschneck, and Gabor Fichtinger.
\newblock {Spinal Curvature Measurement by Tracked Ultrasound Snapshots}.
\newblock {\em Ultrasound in Medicine {\&} Biology}, 40(2):447--454, 2 2014.

\bibitem{Majidi2011AElectronics}
C.~Majidi, R.~Kramer, and R.~J. Wood.
\newblock {A non-differential elastomer curvature sensor for softer-than-skin electronics}.
\newblock {\em Smart Materials and Structures}, 20(10):105017, 8 2011.

\bibitem{Ward-Cherrier2018TheMorphologies}
Benjamin Ward-Cherrier, Nicholas Pestell, Luke Cramphorn, Benjamin Winstone, Maria~Elena Giannaccini, Jonathan Rossiter, and Nathan~F. Lepora.
\newblock {The TacTip Family: Soft Optical Tactile Sensors with 3D-Printed Biomimetic Morphologies}.
\newblock {\em Soft Robotics}, 5(2):216--227, 4 2018.

\bibitem{Li2012SimpleMeasurement}
S.~Li, Q.~He, Z.~Zhang, B.~Han, Z.~Li, and J.~Lan.
\newblock {Simple eddy current sensor for small angle measurement}.
\newblock {\em CPEM Digest (Conference on Precision Electromagnetic Measurements)}, pages 496--497, 2012.

\bibitem{Mineo2017FlexibleComponents}
Carmelo Mineo, Charles MacLeod, Maxim Morozov, S.~Gareth Pierce, Rahul Summan, Tony Rodden, Danial Kahani, Jonathan Powell, Paul McCubbin, Coreen McCubbin, Gavin Munro, Scott Paton, and David Watson.
\newblock {Flexible integration of robotics, ultrasonics and metrology for the inspection of aerospace components}.
\newblock {\em AIP Conference Proceedings}, 1806(1):020026, 2 2017.

\bibitem{Zhang2014FrequencyTransfer}
Yiming Zhang and Zhengming Zhao.
\newblock {Frequency splitting analysis of two-coil resonant wireless power transfer}.
\newblock {\em IEEE Antennas and Wireless Propagation Letters}, 13:400--402, 2014.

\bibitem{Zhu2022}
Peng~Wen Zhu, Xiang Wang, Wen~Sheng Zhao, Jing Wang, Da~Wei Wang, Fang Hou, and Gaofeng Wang.
\newblock Design of h-shaped planar displacement microwave sensors with wide dynamic range.
\newblock {\em Sensors and Actuators A: Physical}, 333:113311, 1 2022.

\bibitem{Horestani2014}
Ali~K. Horestani, Jordi Naqui, Zahra Shaterian, Derek Abbott, Christophe Fumeaux, and Ferran Martín.
\newblock Two-dimensional alignment and displacement sensor based on movable broadside-coupled split ring resonators.
\newblock {\em Sensors and Actuators A: Physical}, 210:18--24, 4 2014.

\bibitem{Babu2016}
Anish Babu and Boby George.
\newblock A linear and high sensitive interfacing scheme for wireless passive lc sensors.
\newblock {\em IEEE Sensors Journal}, 16:8608--8616, 12 2016.

\bibitem{Babu2018}
Anish Babu and Boby George.
\newblock An efficient readout scheme for simultaneous measurement from multiple wireless passive $lc$ sensors.
\newblock {\em IEEE Transactions on Instrumentation and Measurement}, 67:1161--1168, 5 2018.

\bibitem{Babu2019}
Anish Babu and Boby George.
\newblock Sensor system to aid the vehicle alignment for inductive ev chargers.
\newblock {\em IEEE Transactions on Industrial Electronics}, 66:7338--7346, 9 2019.

\bibitem{Jiao2019}
Dian Jiao, Liwei Ni, Xiaoliang Zhu, Jiang Zhe, Ziyu Zhao, Yaguo Lyu, and Zhenxia Liu.
\newblock Measuring gaps using planar inductive sensors based on calculating mutual inductance.
\newblock {\em Sensors and Actuators A: Physical}, 295:59--69, 8 2019.

\bibitem{Kalhor1990TheBiot-Savart}
Hassan~A. Kalhor.
\newblock {The Degree of Intelligence of the Law of Biot-Savart}.
\newblock {\em IEEE Transactions on Education}, 33(4):365--366, 1990.

\bibitem{Zhang2014Frequency-splittingTransfer}
Yiming Zhang, Zhengming Zhao, and Kainan Chen.
\newblock {Frequency-splitting analysis of four-coil resonant wireless power transfer}.
\newblock {\em IEEE Transactions on Industry Applications}, 50(4):2436--2445, 2014.

\bibitem{Hughes2016High-SensitivitySuperalloys}
Robert Hughes.
\newblock {High-Sensitivity Eddy-Current Testing Technology for Defect Detection in Aerospace Superalloys}.
\newblock {\em University of Warwick: Department of Physics}, 2016.

\bibitem{Hughes2014NearDetection}
R.~Hughes, Y.~Fan, and S.~Dixon.
\newblock {Near electrical resonance signal enhancement (NERSE) in eddy-current crack detection}.
\newblock {\em NDT and E International}, 66:82--89, 2014.

\bibitem{Mazlouman2009Mid-rangeSensors}
Shahrzad~Jalali Mazlouman, Alireza Mahanfar, and Bozena Kaminska.
\newblock {Mid-range wireless energy transfer using inductive resonance for wireless sensors}.
\newblock {\em Proceedings - IEEE International Conference on Computer Design: VLSI in Computers and Processors}, pages 517--522, 2009.

\bibitem{Moghaddami2018SensorlessControllers}
Masood Moghaddami, Aditya Sundararajan, and Arif~I. Sarwat.
\newblock {Sensorless electric vehicle detection in inductive charging stations using self-tuning controllers}.
\newblock {\em 2017 IEEE Transportation Electrification Conference, ITEC-India 2017}, 2018-January:1--4, 4 2018.

\bibitem{Hughes2023AnalyticalCoils}
Robert~R Hughes, Alexis Hernandez, and Anthony~J Mulholland.
\newblock {Analytical approximations for magnetic coupling coefficients between adjacent coils}.
\newblock {\em arXiv preprint arXiv:2306.16745}, 2023.

\bibitem{Tyurnev2010CouplingTheory}
V.~V. Tyurnev.
\newblock {Coupling Coefficients of Resonators in Microwave Filter Theory}.
\newblock {\em Progress In Electromagnetics Research B}, 21(21):47--67, 2010.

\bibitem{Tung2005ACoupling}
Wei~Shin Tung, Yi~Chyun Chiang, and Jui~Ching Cheng.
\newblock {A new compact LTCC bandpass filter using negative coupling}.
\newblock {\em IEEE Microwave and Wireless Components Letters}, 15(10):641--643, 10 2005.

\bibitem{Petrov2015RetrievalCoefficients}
P.~Petrov, A.~Radkovskaya, and E.~Shamonina.
\newblock {Retrieval of electric and magnetic coupling coefficients}.
\newblock {\em 2015 9th International Congress on Advanced Electromagnetic Materials in Microwaves and Optics, METAMATERIALS 2015}, pages 259--261, 11 2015.

\bibitem{Wang2018AnTransfer}
Yijie Wang, Yousu Yao, Xiaosheng Liu, Dianguo Xu, and Liang Cai.
\newblock {An LC/S Compensation Topology and Coil Design Technique for Wireless Power Transfer}.
\newblock {\em IEEE Transactions on Power Electronics}, 33(3):2007--2025, 3 2018.

\bibitem{Zhang2015SelectiveFrequencies}
Yiming Zhang, Ting Lu, Zhengming Zhao, Fanbo He, Kainan Chen, and Liqiang Yuan.
\newblock {Selective Wireless Power Transfer to Multiple Loads Using Receivers of Different Resonant Frequencies}.
\newblock {\em IEEE Transactions on Power Electronics}, 30(11):6001--6005, 11 2015.

\bibitem{Yang2017DesignCoils}
Dongsheng Yang, Sokhui Won, and Huan Hong.
\newblock {Design of Range Adaptive Wireless Power Transfer System Using Non-coaxial Coils}.
\newblock {\em IOP Conference Series: Materials Science and Engineering}, 199(1):012008, 5 2017.

\end{thebibliography}

\end{document}